\documentclass[12pt,a4paper]{article}

\usepackage[utf8]{inputenc}
\usepackage[T1]{fontenc}
\usepackage{amsmath,amssymb}
\usepackage{booktabs}
\usepackage{longtable}
\usepackage{array}
\usepackage{graphicx}
\usepackage{hyperref}
\usepackage[margin=1in]{geometry}
\usepackage{enumitem}
\usepackage{microtype}
\usepackage{setspace}
\usepackage{tikz}
\usetikzlibrary{arrows.meta,positioning,shapes.geometric}

\title{Regulatory Expectations for Bayesian Methods in Drug and Biologic Clinical Trials:\\ A Practical Perspective on FDA's 2026 Draft Guidance}
\author{Yuan Ji, PhD}
\date{January 2026}

\begin{document}

\maketitle

\begin{abstract}
The U.S.\ Food and Drug Administration (FDA) released a landmark draft guidance in January 2026 on the use of Bayesian methodology to support primary inference in clinical trials of drugs and biological products~\cite{ref1, ref2, ref3}. For sponsors, the central message is not merely that ``Bayes is allowed,'' but that Bayesian designs should be justified through explicit success criteria, thoughtful priors (especially when borrowing external information), prospective operating-characteristic evaluation (often via simulation when simulation is used), and computational transparency suitable for regulatory review~\cite{ref2}. This paper provides a practical, regulatory-oriented synthesis of the draft guidance, highlighting where Bayesian designs can be calibrated to traditional frequentist error-rate targets and where, with sponsor--FDA agreement, alternative Bayesian operating metrics may be appropriate~\cite{ref2, ref4}. We illustrate expectations through examples discussed in the guidance (e.g., platform trials, external/nonconcurrent controls, pediatric extrapolation) and conclude with an actionable checklist for planning documents and submission packages.
\end{abstract}

\begin{figure}[htbp]
\centering
\includegraphics[width=\textwidth]{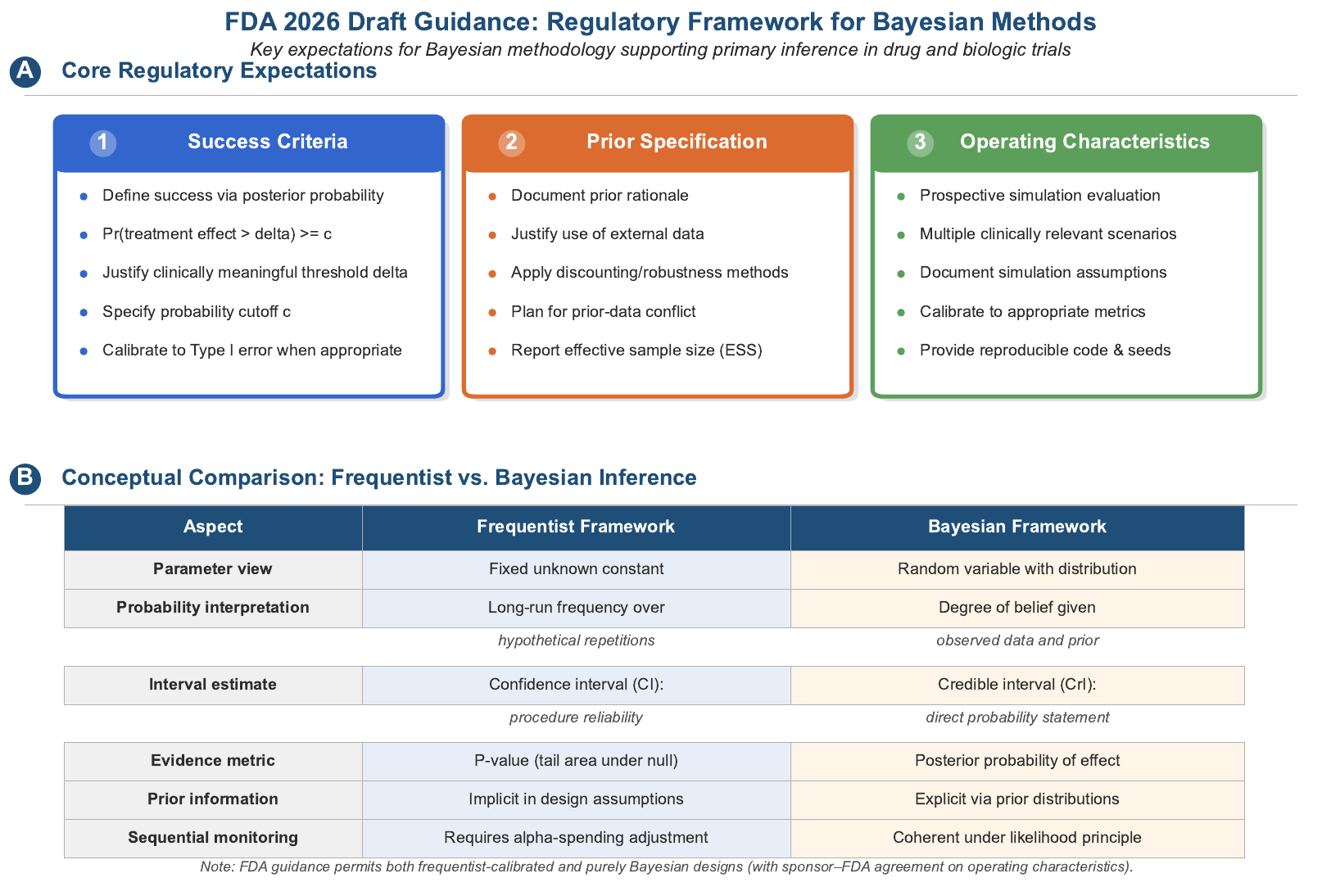}
\caption{Overview of FDA's 2026 draft guidance on Bayesian methodology in clinical trials. Panel~A summarizes the three core regulatory expectations: (1)~success criteria defined via posterior probabilities, (2)~principled prior specification with safeguards for borrowing, and (3)~prospective evaluation of operating characteristics. Panel~B contrasts key conceptual differences between frequentist and Bayesian inferential frameworks.}
\label{fig:overview}
\end{figure}

\section{Introduction and Regulatory Context}

Bayesian methods are increasingly used across the clinical development lifecycle, including early-phase dose finding, platform trials, and confirmatory settings where borrowing external information may improve efficiency~\cite{ref5, ref6}. The January 2026 FDA draft guidance provides a unifying regulatory framework for how such methods should be specified, justified, and documented when used to support primary inference for drugs and biological products~\cite{ref1, ref2}. Throughout, the guidance emphasizes that Bayesian flexibility comes with a higher bar for up-front design evaluation and transparent, reproducible analysis implementation~\cite{ref2}.

\subsection{Why Bayesian methods? Practical benefits}

From a sponsor and regulator standpoint, the benefits of Bayesian methods are primarily practical: they enable \emph{direct probability statements} about clinically meaningful effects, support \emph{explicit decision criteria} (e.g., posterior-probability or predictive-probability rules), and provide a coherent framework for \emph{integrating prior/external information} when justified (with safeguards against prior--data conflict)~\cite{ref6, ref2, ref7}. These features are especially valuable in settings where conventional designs can be inefficient or infeasible (e.g., rare diseases, pediatrics, and platform trials), but they also shift attention toward careful up-front evaluation of operating characteristics and robust documentation~\cite{ref2}.

An important practical claim---and one supported by a substantial methodological literature---is that Bayesian designs can be \emph{more efficient} than conventional frequentist designs under comparable evidentiary goals. Here ``efficiency'' typically refers to reductions in expected sample size or trial duration (via early stopping for efficacy or futility), improved decision quality under uncertainty, or improved information use through borrowing, while maintaining acceptable operating characteristics under scenarios relevant to regulators~\cite{ref6, ref7, ref2}. For example, comparative studies of Bayesian sequential monitoring (using posterior or predictive probabilities) and frequentist group sequential designs show that, for appropriately calibrated thresholds, Bayesian designs can achieve similar frequentist error control and power with favorable expected sample size profiles in many scenarios~\cite{ref8, ref4}. More broadly, reviews of Bayesian adaptive trial methods emphasize that pre-specified Bayesian decision rules---combined with simulation-based operating-characteristic evaluation---can yield meaningful resource savings and faster learning, particularly in settings where accumulating information during the trial is leveraged for adaptive decisions~\cite{ref5, ref9}.

\subsection{A brief primer: Bayes' theorem and posterior updating}

Bayesian inference combines a prior distribution \(p(\theta)\) for unknown parameters \(\theta\) with a likelihood \(p(D\mid \theta)\) for observed data \(D\), yielding a posterior distribution \(p(\theta\mid D)\) via Bayes' theorem:
\begin{equation}
p(\theta \mid D) = \frac{p(D \mid \theta)\,p(\theta)}{\int p(D \mid \theta)\,p(\theta)\,d\theta}.
\end{equation}
This posterior distribution is the basis for reporting credible intervals and computing quantities used in Bayesian decision-making (e.g., posterior probabilities of clinically relevant effect thresholds, or predictive probabilities of trial success)~\cite{ref7, ref2}.

Bayesian updating is particularly well suited to drug development because programs routinely accumulate \emph{structured prior information} before a new confirmatory study starts. For example, suppose \(\theta\) is a control-arm response rate for a given indication and endpoint. Historical data from earlier phases in the same program, prior studies of the drug class, or high-quality external evidence can be encoded as a prior distribution for \(\theta\) (with robustness/sensitivity analyses to address potential non-exchangeability). As new trial data accrue, the posterior distribution provides an updated, quantitative summary that integrates what was known before the study with what the study has observed, while making the assumptions explicit and reviewable~\cite{ref6, ref7, ref2}.

In simple conjugate settings, this updating is transparent. For a binary endpoint with a Beta prior \(\theta\sim\text{Beta}(a_0,b_0)\) and observed data \(y\) responses out of \(n\) patients, the posterior is \(\theta\mid D\sim\text{Beta}(a_0+y,\;b_0+n-y)\). Here the prior parameters \((a_0,b_0)\) can be interpreted as ``pseudo-counts'' reflecting the strength of prior information (e.g., from historical controls), which can be discounted or made robust when external information may not fully align with the current population~\cite{ref10, ref2}.
In this Beta--Binomial setting, the quantity \((a_0+b_0)\) is often interpreted as the \emph{prior effective sample size} (ESS), i.e., the amount of information contributed by the prior in units comparable to binomial observations~\cite{ref11}.

Bayesian summaries often align with how clinical and operational decisions are actually made under uncertainty: decision-makers typically want probabilities of clinically meaningful statements (e.g., ``the probability the treatment effect exceeds a clinically relevant threshold'') rather than tail-area measures under a sharp null. A recurring pain point in drug development is the gap between \emph{statistical significance} and \emph{clinical significance}: a small $p$-value can occur for a trivial effect in a large sample, while a clinically important effect can fail to achieve ``statistical significance'' in small or heterogeneous populations. This concern has been highlighted in the ASA-sponsored 2019 \emph{The American Statistician} special issue calling for a ``world beyond $p<0.05$'' and discouraging dichotomous ``significant/non-significant'' declarations, as well as in Nature commentary arguing that the term ``statistical significance'' should be abandoned~\cite{ref12, ref13}.

Finally, Bayesian success criteria expressed as posterior probabilities naturally address clinical relevance by explicitly incorporating an effect threshold \(a\) (e.g., a minimum clinically important difference) and then requiring high posterior certainty, such as \(\Pr(d>a\mid \text{data}) \ge c\). The FDA draft guidance explicitly discusses posterior-probability success criteria and emphasizes that, regardless of whether designs are calibrated to frequentist Type I error rate, conclusions must be supported by scenario-based operating characteristics and sensitivity analyses that reflect uncertainties in priors, models, and data quality~\cite{ref2}.

\section{Key inferential distinctions that matter for regulatory review}

The practical differences between Bayesian and frequentist frameworks manifest in the measures used for hypothesis testing and interval estimation. At a foundational level, Bayesian statistics assigns two probability measures: one to the data model \(p(D\mid \theta)\) and another to the unknown parameters \(p(\theta)\). Frequentist statistics assigns a probability model to the data, but treats parameters as fixed unknown constants rather than random variables with probability distributions. As a result, inference and decision-making in the two paradigms are conceptually different, and frequentist inference cannot directly answer many scientific questions in the form of a probability statement about the treatment effect (e.g., ``the probability that \(d\) exceeds a clinically meaningful threshold'') without additional assumptions or reinterpretation. While frequentist models focus on rejecting a null hypothesis, Bayesian models emphasize the estimation of a treatment's effect size and the uncertainty surrounding it~\cite{ref14, ref15}.

\begin{table}[htbp]
\centering
\caption{Comparison of Frequentist and Bayesian Mechanisms}
\label{tab:comparison}
\begin{tabular}{@{}p{2.5cm}p{4cm}p{4cm}p{4cm}@{}}
\toprule
\textbf{Feature} & \textbf{Frequentist Mechanism} & \textbf{Bayesian Mechanism} & \textbf{Clinical Implication} \\
\midrule
Probability Source & Long-run frequency in infinite repetitions & Subjective degree of belief updated by data & Allows for direct statements about hypothesis truth~\cite{ref14} \\
\addlinespace
Parameter View & Fixed, unknown constant & Random quantity with a probability distribution & Quantifies ``how likely'' a specific effect size is~\cite{ref15} \\
\addlinespace
Interval Metric & 95\% Confidence Interval (CI) & 95\% Credible Interval (CrI) & CrI is the probability that the parameter is in the range~\cite{ref14} \\
\addlinespace
Evidence Metric & $P$-value (extremity of data) & Posterior probability (likelihood of effect) & Bayesian metrics are more intuitive for decision-making~\cite{ref14} \\
\addlinespace
Handling Prior Data & Typically implicit (used for planning assumptions and design inputs) & Formal (integrated via prior distributions) & Efficiency gains by leveraging existing knowledge when assumptions are explicit and stress-tested~\cite{ref6} \\
\bottomrule
\end{tabular}
\end{table}

The frequentist 95\% CI is a statement about the reliability of the estimation procedure; it suggests that if the experiment were repeated many times, 95\% of the intervals would contain the true parameter. It does not mean there is a 95\% chance the parameter is in the current interval. The Bayesian 95\% CrI, however, provides exactly that: a direct probabilistic statement that the parameter resides within the specified bounds given the observed data and prior information~\cite{ref14}.

Importantly, it is a misconception that frequentist practice is ``prior-free.'' While frequentist inference does not assign a probability distribution to an unknown fixed parameter, it routinely relies on \emph{prior information implicitly} through the choice of design assumptions and tuning constants. For example, standard frequentist sample size and power calculations require specifying (i) nuisance parameters such as the control-arm event rate or outcome variance and (ii) a target effect size under an alternative hypothesis; these inputs are almost never known a priori and are typically informed by previous studies, pilot data, registries, or expert clinical judgment~\cite{ref16}. In this sense, the design is anchored to prior beliefs, but those beliefs are not mathematically represented and propagated as uncertainty. Bayesian designs make the same dependency explicit by writing a prior distribution for the parameters, which can then be debated, stress-tested, and updated transparently as new evidence accrues~\cite{ref17, ref18}. This explicitness is often viewed as a feature rather than a bug: disagreements about priors are surfaced and can be evaluated via sensitivity analyses, whereas implicit assumptions in frequentist planning can be harder to identify and quantify.

\section{Core expectations in the 2026 FDA draft guidance}

A landmark development in this field occurred on January 12, 2026, when FDA released a draft guidance titled ``Use of Bayesian Methodology in Clinical Trials of Drug and Biological Products''~\cite{ref1, ref2, ref3}. Issued jointly by CDER and CBER, the draft guidance signals a formal regulatory shift toward the acceptance of Bayesian primary inference in pivotal confirmatory trials, provided designs are justified through success criteria, operating characteristics, and robust documentation~\cite{ref2}.

 The FDA's guidance emphasizes that Bayesian methodologies can help address persistent hurdles in drug development, including cost and timeline constraints, particularly in settings where conventional designs are impractical~\cite{ref2, ref19}. The document is also notable for translating Bayesian concepts into operational expectations that sponsors can use to justify designs and analyses in submissions. Four practical takeaways are presented next to summarize a few key points from the guidance~\cite{ref1}.

\begin{figure}[htbp]
\centering
\begin{tikzpicture}[
  node distance=6mm and 6mm,
  box/.style={draw, rounded corners, align=center, text width=0.23\textwidth, minimum height=10mm, font=\small},
  arrow/.style={-Latex, thick}
]
\node[box] (estimand) {Define estimand\\\& success rule\\(posterior threshold; clinical $a$)};
\node[box, right=of estimand] (priors) {Specify priors\\Analysis \& design prior\\Borrowing/discounting};
\node[box, right=of priors] (ocs) {Operating characteristics\\Scenario grid; simulation\\Frequentist or Bayesian OCs};

\node[box, below=of priors] (compute) {Computation\\Software/version; code\\Algorithm settings; seeds};
\node[box, left=of compute] (protocol) {Protocol/SAP\\Decision rules\\Interim schedule};
\node[box, right=of compute] (submission) {Submission package\\Simulation report\\Sensitivity analyses};

\draw[arrow] (estimand) -- (priors);
\draw[arrow] (priors) -- (ocs);
\draw[arrow] (ocs) -- (submission);
\draw[arrow] (priors) -- (compute);
\draw[arrow] (compute) -- (submission);
\draw[arrow] (estimand) -- (protocol);
\draw[arrow] (protocol) -- (priors);
\end{tikzpicture}
\caption{Sponsor-facing workflow implied by FDA's 2026 draft guidance when Bayesian methods support primary inference.}
\label{fig:workflow}
\end{figure}
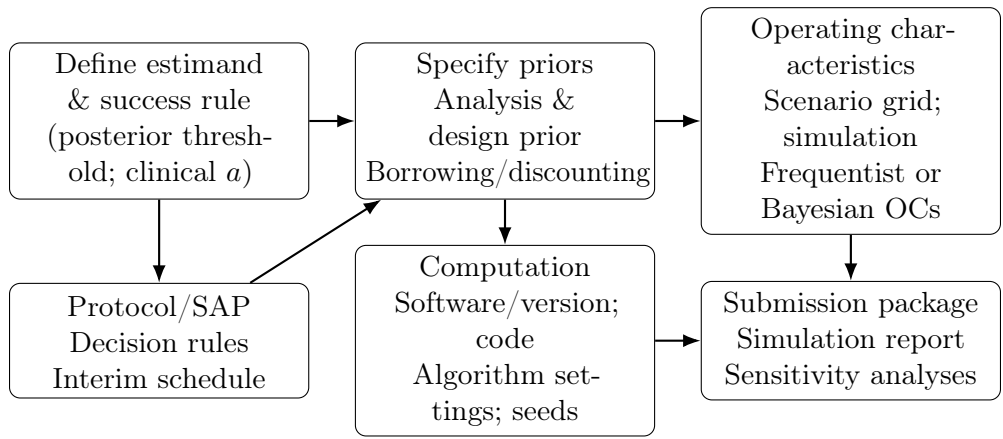

Figure~\ref{fig:workflow} provides a sponsor-facing schematic of the end-to-end workflow implied by the draft guidance when Bayesian methods support primary inference.

\subsection{Four practical takeaways from the 2026 draft guidance}

\paragraph{1) Success criteria can be defined in posterior-probability terms.}
 Rather than defining success solely through a frequentist $p$-value, the guidance discusses Bayesian success criteria expressed through posterior probabilities, e.g., \(\Pr(d>a)>c\), where \(d\) is a population-level treatment effect summary, \(a\) is a clinically meaningful threshold, and \(c\) is a required posterior probability level~\cite{ref2}. Importantly, the guidance recognizes multiple paradigms for choosing these thresholds: in some settings, \(a\) and \(c\) can be selected to control frequentist error rates (e.g., overall FWER), while in other settings---when sponsor and FDA agree Type I error rate calibration is not required---design calibration may instead target Bayesian operating metrics defined with respect to a design (sampling) prior, such as Bayesian power/assurance and probability of a correct decision~\cite{ref2, ref7}. The exact impact of calibrating a Bayesian decision rule to controlling the frequentist error rate is not clear and might require further investigation. This concern has also been raised forcefully in methodological commentary by Frank Harrell, who argues that ``wedding'' Bayesian decision-making to frequentist error-rate constraints can create conceptual and practical confusion, and that the probabilities decision-makers need are forward-time (posterior) probabilities rather than long-run Type I error rates~\cite{ref20, ref21}.

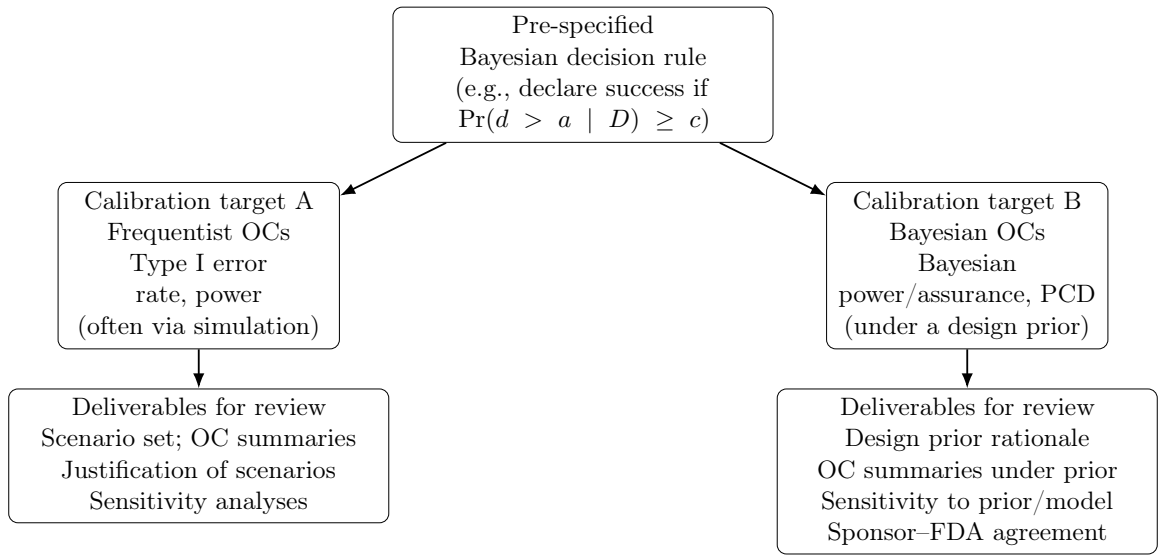
\begin{figure}[htbp]
\centering
\resizebox{0.95\textwidth}{!}{%
\begin{tikzpicture}[
  node distance=6mm and 8mm,
  box/.style={draw, rounded corners, align=center, text width=5.5cm, minimum height=10mm, font=\small},
  small/.style={draw, rounded corners, align=center, text width=4cm, minimum height=9mm, font=\small},
  arrow/.style={-Latex, thick}
]
\node[box] (rule) {Pre-specified Bayesian decision rule\\(e.g., declare success if $\Pr(d>a\mid D)\ge c$)};

\node[small, below left=of rule] (calibF) {Calibration target A\\Frequentist OCs\\Type I error rate, power\\(often via simulation)};
\node[small, below right=of rule] (calibB) {Calibration target B\\Bayesian OCs\\Bayesian power/assurance, PCD\\(under a design prior)};

\node[box, below=of calibF] (deliverF) {Deliverables for review\\Scenario set; OC summaries\\Justification of scenarios\\Sensitivity analyses};
\node[box, below=of calibB] (deliverB) {Deliverables for review\\Design prior rationale\\OC summaries under prior\\Sensitivity to prior/model\\Sponsor--FDA agreement};

\draw[arrow] (rule) -- (calibF);
\draw[arrow] (rule) -- (calibB);
\draw[arrow] (calibF) -- (deliverF);
\draw[arrow] (calibB) -- (deliverB);
\end{tikzpicture}%
}
\caption{Two common calibration targets for posterior-/predictive-probability decision rules: calibration to frequentist operating characteristics (OCs) versus calibration to Bayesian operating metrics (with sponsor--FDA agreement in specific contexts).}
\label{fig:calibration}
\end{figure}

\paragraph{2) Borrowing external information is feasible, but requires safeguards.}
The draft guidance formalizes and contextualizes the use of informative priors to borrow external evidence (e.g., previous trials, external controls, adult data for pediatric extrapolation) as a strategy to improve efficiency in settings where fully powered randomized trials are infeasible~\cite{ref2}. At the same time, it emphasizes that borrowing must be justified and pre-specified, and that sponsors should incorporate safeguards against prior-data conflict through approaches such as discounting and sensitivity analyses~\cite{ref2, ref22}.
In addition to rare diseases and pediatrics, precision medicine is likely to make this motivation more common: as therapies are targeted to increasingly narrow biomarker-defined subpopulations, target populations (and thus feasible sample sizes per subgroup) tend to shrink, making standalone, fully powered trials harder to conduct for every clinically relevant stratum~\cite{ref23, ref5}. In such settings, principled information sharing---across related subgroups, across trial stages within a development program, or from high-quality external data---may become increasingly necessary to support timely and reliable regulatory decision making, provided the exchangeability assumptions are justified and the resulting operating characteristics are thoroughly evaluated~\cite{ref2, ref6}.

\paragraph{3) Operating characteristics should be evaluated, often via simulation.}
Because Bayesian operating characteristics depend on the prior and the data-generating assumptions, the guidance emphasizes careful, prospective evaluation of performance across scenarios that are relevant to the intended use of the trial results~\cite{ref2}. When simulations are used to estimate operating characteristics (e.g., Type I error rate/power for calibrated designs or Bayesian power/probability of correct decision for non-calibrated designs), sponsors should provide clear documentation of the simulation design, assumptions, and code to support verification and review~\cite{ref2}.

\paragraph{4) Reproducibility and computational transparency are elevated.}
Finally, the guidance raises expectations for computational reliability and transparency. It emphasizes reliable software, documentation of computational procedures (including MCMC settings and convergence diagnostics when applicable), software/version information, and providing documented code for primary and key secondary analyses and sensitivity analyses; for MCMC analyses, the guidance discusses reporting details (e.g., seed number(s) used during chain initiation) to support reproducibility~\cite{ref2}.

\section{Taxonomy of Bayesian Clinical Trial Designs}

Bayesian methodology is not a monolithic entity but a collection of diverse designs tailored to different phases of the clinical development lifecycle. Each design leverages the sequential nature of Bayesian updating to achieve specific goals, from safety signals in early phases to efficacy confirmation in late phases~\cite{ref24, ref25}.

\subsection{Phase I: Model-based, model-assisted, model-free, and hybrid designs}

Phase I trials represent the most successful application of Bayesian methods to date. The traditional ``3+3'' design, while simple, is statistically inefficient and often exposes patients to sub-therapeutic or overly toxic doses. The Continual Reassessment Method (CRM), pioneered by O'Quigley, Pepe, and Fisher in 1990, replaced algorithm-based dosing with model-based estimation~\cite{ref26}.

The CRM uses a Bayesian framework to estimate the probability of dose-limiting toxicity (DLT) across several dose levels. As each patient's outcome is observed, the model updates the posterior distribution of the DLT probability, and the next patient is assigned to the dose level currently estimated to be closest to the target toxicity rate (e.g., 20\% or 30\%)~\cite{ref26}.

\paragraph{Evolution of CRM and Likelihood Alternatives}

Subsequent refinements have addressed early-stage stability and complexity:

\begin{itemize}[leftmargin=*]
    \item \textbf{Likelihood CRM:} Although CRM is often discussed under a Bayesian implementation, the original CRM is not Bayesian \emph{per se} in the sense that it can be implemented without an explicit prior; later Bayesian versions place a prior distribution on the parameters of the dose--toxicity curve and update that prior as DLT outcomes accrue. O'Quigley and Shen (1996) demonstrated that CRM can also function using a likelihood-based framework, though it requires at least one toxicity to be observed before the model can ``get off the ground.'' This often necessitates a hybrid approach where the trial starts with a standard Up-and-Down scheme before switching to the model~\cite{ref27}.

    \item \textbf{mTPI and mTPI-2:} The modified toxicity probability interval (mTPI) design and its successor mTPI-2 are Bayesian, curve-free (model-assisted) dose-finding designs that use simple Beta--Binomial modeling for estimating toxicity probabilities while retaining transparent, pre-tabulated escalation/de-escalation rules based on posterior probabilities for under-, target-, and over-toxicity intervals around a target DLT rate~\cite{ref28, ref29}. These designs pioneered interval decision rules as a principled form of up-and-down rules, helping open a new chapter of practical, interpretable dose-finding methodologies.

    \item \textbf{BOIN (Bayesian Optimal INterval):} BOIN is a model-assisted interval design with pre-specified escalation/de-escalation boundaries, which makes implementation straightforward and transparent in practice~\cite{ref10, ref30}. Its operating characteristics depend on the target DLT rate and the chosen boundaries (including any additional practical constraints on escalation/de-escalation). In small cohorts, certain observation patterns (e.g., 1/3 or 2/3 DLTs) can lead to unexpected or repeated de-escalation decisions if the same dose is revisited, so protocols often pre-specify rules to limit oscillation and sponsors typically evaluate these behaviors via simulation under clinically plausible scenarios.

    \item \textbf{Drug Combinations:} In dual-agent trials, the assumption of toxicity monotonicity (that higher doses always mean higher toxicity) often fails. Bayesian hierarchical models allow for partial ordering, enabling the exploration of complex dose-response surfaces~\cite{ref31}.
\end{itemize}

Model-free (algorithmic) designs remain widely used in practice. The classical \textbf{3+3} design is a cohort-based escalation rule that determines dose escalation/de-escalation based only on the number of DLTs observed in small cohorts, without an explicit statistical model. A more recent refinement, the \textbf{i3+3} (interval 3+3) design retains the operational simplicity of 3+3 but introduces an equivalence interval around a target toxicity rate to make more stable escalation decisions~\cite{ref32}. Hybrid approaches that blend algorithmic rules with model-based components have also been proposed; for example, the design in~\cite{ref33} provides a hybrid framework intended to preserve practical simplicity while improving operating characteristics.

\subsection{Phase II and III: Bayesian designs for adaptive decision rules and sequential monitoring}

Bayesian  designs allow for modifications to the trial based on accumulating data without the heavy ``multiplicity penalties'' associated with frequentist group sequential designs~\cite{ref34, ref35}. These modifications include early stopping for efficacy or futility, adjusting the randomization ratio to favor more effective treatments (Response-Adaptive Randomization), or enriching the trial population with patients most likely to respond~\cite{ref35, ref9}. I discuss a few key points next.

\paragraph{Multiplicity, repeated looks, and what Bayesians mean by ``multiple testing.''}
A key conceptual distinction in Bayesian trial design is between (i) repeatedly evaluating the \emph{same} scientific question as data accrue (e.g., interim monitoring for efficacy/futility for a single primary estimand), and (ii) conducting \emph{multiple different} tests/questions (e.g., multiple endpoints, multiple subgroups, or multiple treatment comparisons). From a Bayesian perspective grounded in the likelihood principle, posterior probabilities remain coherent under optional stopping: if the model and prior are fixed in advance, repeated interim looks do not inherently require the same ``multiplicity adjustment'' logic as frequentist repeated significance testing because Bayesian inference conditions on the data actually observed rather than on a sampling plan over hypothetical data~\cite{ref4, ref36}. In contrast, when a program makes multiple distinct claims across endpoints or subpopulations, Bayesian modeling often addresses the multiplicity of scientific questions by explicitly representing relationships across parameters (e.g., hierarchical models, shrinkage, or partial pooling), and by pre-specifying decision criteria for each claim and evaluating the joint operating characteristics of the overall decision strategy~\cite{ref2, ref37}.

\paragraph{Bayesian group sequential designs: posterior and predictive probability boundaries.}
In Phase II/III settings, Bayesian group sequential designs typically use stopping/decision rules based on posterior probabilities (PP) or posterior predictive probabilities of success (PPOS). For example, a trial may stop early for efficacy if the posterior probability that the treatment effect exceeds a clinically meaningful threshold is sufficiently high, or stop for futility if the predictive probability of ultimately meeting a success criterion is sufficiently low. When such rules are prespecified, they can support efficient decision-making and ethical conduct by allowing earlier conclusions under strong evidence, while still requiring thorough pre-trial simulation to characterize performance across clinically relevant scenarios~\cite{ref2, ref8, ref4, ref36}.

\paragraph{The Type I error rate calibration debate: pragmatism versus principle.}
The FDA draft guidance discusses calibrating Bayesian decision rules to frequentist Type I error rate in some settings, reflecting a regulatory interest in long-run false positive rates under null scenarios~\cite{ref2}. However, there is an important internal debate among Bayesians about this practice. From a strict likelihood-principle (``subjective Bayesian'') viewpoint, calibration to a frequentist Type I error rate can be seen as in tension with purely Bayesian evidence reporting. One critique is that frequentist Type I error rate is typically computed under a point null (a fixed ``no effect'' data-generating model), whereas Bayesian inference is built around a prior distribution representing uncertainty about effect sizes; treating Type I error rate calibration as if it implied a point-mass prior at the null would be inconsistent with how Bayesian analyses are actually specified in practice. In applied development programs, Bayesian designs typically use proper priors and then \emph{overlay} a frequentist calibration step on posterior- or predictive-probability cutoffs to meet regulatory expectations; critics argue that this overlay can yield conservative decision thresholds (increasing false negatives) and can blur the conceptual distinction between Bayesian evidence measures and frequentist error-rate constraints~\cite{ref20, ref21, ref36, ref38}. The FDA guidance also explicitly allows, with sponsor--FDA agreement in specific contexts, Bayesian designs that are not calibrated to Type I error rate provided operating characteristics are justified using Bayesian metrics and comprehensive scenario evaluation~\cite{ref2}.

\paragraph{Type I error versus Type I error rate.}
It is important to distinguish the \emph{Type I error} from the \emph{Type I error rate}. A Type I error is a specific decision error event (e.g., declaring efficacy when the treatment is not efficacious under the chosen null scenario). The Type I error \emph{rate} is a frequentist long-run probability: it is the probability of committing a Type I error, computed over repeated hypothetical realizations of the trial under a prespecified null data-generating mechanism (often a point null effect size), together with the full decision procedure (including interim looks and stopping rules). In this sense, the Type I error rate is a property of the overall design-and-decision rule under repetition, whereas Bayesian posterior probabilities are conditional on the observed data and the specified prior and likelihood.

\section{Information Borrowing via Bayesian Hierarchical Models}

The most distinctive capability of Bayesian trials is the formal integration of external data---a process known as information borrowing. This is primarily achieved through Bayesian hierarchical models, which assume that treatment effects across studies or subgroups are not identical but may be modeled as exchangeable draws from a common distribution; the degree of borrowing is then determined by both the model structure and the observed consistency between sources~\cite{ref37, ref10}.

\paragraph{Mechanisms of Informative Priors}

Borrowing strength from external sources requires the construction of an informative prior. The degree of ``borrowing'' is often determined by the consistency between the historical data and the current trial data.

\begin{table}[htbp]
\centering
\caption{Mechanisms of Informative Priors}
\label{tab:priors}
\begin{tabular}{@{}p{2.5cm}p{5cm}p{5cm}@{}}
\toprule
\textbf{Borrowing Method} & \textbf{Mechanism} & \textbf{Risk/Benefit} \\
\midrule
Power Priors & Down-weights historical data using a parameter $\alpha_0$ & Protects against historical data dominance~\cite{ref10} \\
\addlinespace
MAP Priors & Synthesizes multiple historical trials into a predictive distribution & Highly efficient for stable historical data~\cite{ref10, ref31} \\
\addlinespace
Robust MAP & Uses a mixture prior (MAP + non-informative component) & Protects against ``prior-data conflict''~\cite{ref14, ref10} \\
\addlinespace
Commensurate Priors & Hierarchical prior that links historical and current parameters, with borrowing determined by commensurability between data sources & Dynamic borrowing that can down-weight external information under prior--data conflict~\cite{ref2} \\
\bottomrule
\end{tabular}
\end{table}

A critical insight from the literature is that when historical results differ substantially from current data, principled hierarchical/robust formulations can reduce borrowing and increase uncertainty, helping mitigate bias from overly optimistic external information~\cite{ref10, ref14}.

\section{Case Study Analysis: I-SPY 2 and Platform Innovations}

The I-SPY 2 trial (Investigation of Serial Studies to Predict Your Therapeutic Response through Imaging and Molecular Analysis 2) serves as the definitive landmark for Bayesian platform trials. Designed to evaluate multiple investigational agents in the neoadjuvant breast cancer setting, it utilized pathologic complete response (pCR) as a surrogate endpoint to accelerate drug graduation to Phase III~\cite{ref39, ref40}.

\subsection{The ``Time Machine'' and Molecular Signatures}

The I-SPY 2 platform implemented several groundbreaking Bayesian techniques:

\begin{itemize}[leftmargin=*]
    \item \textbf{The Time Machine:} Because platform trials run for years, the standard of care or the patient population may shift. The Bayesian ``Time Machine'' methodology allows for bridging across treatment arms and different time periods to inform ongoing analyses while accounting for temporal drift~\cite{ref2, ref41}.

    \item \textbf{Marker-Defined Subgroups:} Patients are classified into one of ten molecular subtypes. Bayesian predictive modeling evaluates 255 possible subtype partitions to find the ``graduation signature'' for each drug~\cite{ref42, ref41}.

    \item \textbf{Predictive Probabilities:} A drug ``graduates'' if there is an 85\% Bayesian predictive probability that it will be successful in a subsequent 300-patient Phase III trial. Conversely, if the probability of success drops below 10\% for all signatures, the drug is dropped for futility~\cite{ref40}.
\end{itemize}

The success of I-SPY 2 led to the graduation of multiple therapies, including pembrolizumab, significantly reducing the time and cost of moving from Phase II to Phase III~\cite{ref42, ref40}.

\subsection{Additional examples highlighted in the FDA draft guidance.}
Beyond I-SPY 2, the FDA draft guidance cites a range of concrete examples to illustrate where Bayesian methods (especially borrowing and platform-trial analyses) are being used or proposed in practice~\cite{ref2}. These include: (i) oncology platform trials such as \emph{GBM AGILE} and \emph{Precision Promise}, where Bayesian analyses have been proposed to augment randomized concurrent controls with external or nonconcurrent control information while accounting for temporal drift; (ii) borrowing from prior studies in a Phase III study of \emph{REBYOTA} (fecal microbiota, live-jslm) for prevention of recurrent \emph{Clostridioides difficile} infection; (iii) pediatric extrapolation and other submissions using Bayesian analyses to leverage historical evidence from reference populations (e.g., active-comparator information) when randomized pediatric trials are challenging; and (iv) borrowing across related diseases/subtypes under a common master protocol (e.g., a proposed epilepsy study in patients with myoclonic-atonic seizures)~\cite{ref2}. Collectively, these examples underscore FDA's emphasis on carefully justified borrowing, scenario-based operating-characteristic evaluation, and transparency about assumptions and potential sources of bias when external/nonconcurrent information is incorporated.

\subsection{Faster Drug Discovery Calls for Platform Trials}

Recent breakthroughs in artificial intelligence and machine learning are accelerating parts of the drug discovery and early development pipeline, increasing the number of plausible candidate therapies that can enter clinical testing~\cite{ref43}. At the same time, modern development programs increasingly target narrower, biomarker-defined patient subpopulations. This combination---more candidates to test but fewer eligible patients per subgroup---creates a fundamental throughput constraint for the clinical enterprise: patient populations are limited by nature, but the number of therapeutically motivated hypotheses is growing.

Platform trials and other master-protocol approaches are therefore likely to become increasingly important to evaluate a growing set of candidate therapies more efficiently. By using a common infrastructure and (often) shared control information, platform trials can add and drop treatment arms over time, learn across interim looks, and allocate limited patient resources to the most informative comparisons, while maintaining a coherent evidentiary framework for decision-making~\cite{ref44}. The FDA draft guidance's emphasis on prospective operating-characteristic evaluation, transparent decision rules, and documentation of assumptions is especially salient for platforms, where time trends, nonconcurrent controls, and adaptation rules can materially affect inference~\cite{ref2}.

In this setting, Bayesian modeling naturally supports information sharing: hierarchical structures can partially pool across related subgroups or arms when justified, and robust borrowing frameworks can leverage external or prior program data with safeguards against prior--data conflict. When done well, these tools can increase precision and improve decision quality in small subgroups, but they also raise the bar for simulation-based evaluation and sensitivity analyses that demonstrate robustness under plausible drift and heterogeneity scenarios---a theme repeatedly emphasized in the draft guidance~\cite{ref2, ref10}.

\section{Operationalizing frequentist calibration in Bayesian sequential designs}

The discussion above highlighted the conceptual and practical tensions around calibrating Bayesian decision rules to a frequentist Type I error rate. Here we focus on an implementation-oriented synthesis that is useful for protocol writers and reviewers. Zhou and Ji~\cite{ref4} emphasize that whether interim looks ``require adjustment'' depends on the inferential perspective (see Table~\ref{tab:stopping}). From a \emph{frequentist-oriented} perspective, one calibrates posterior- or predictive-probability stopping rules to control frequentist operating characteristics (e.g., Type I error rate). From a \emph{calibrated Bayesian} perspective, calibration is treated as a design-stage choice to align Bayesian decision rules with regulatory error-rate targets while retaining Bayesian evidentiary measures. From a \emph{subjective Bayesian} perspective grounded in the likelihood principle, posterior probabilities retain their interpretation regardless of the stopping rule or number of interim analyses; the focus shifts from long-run error rates to coherent probability statements and decision-theoretic trade-offs~\cite{ref4}. This framing connects directly to the FDA draft guidance, which (i) emphasizes simulation-based evaluation of operating characteristics and, when appropriate, calibration of Bayesian decision rules to traditional Type I error rate targets, and (ii) also allows, with sponsor--FDA agreement in specific contexts, designs not calibrated to Type I error rate provided operating characteristics are justified using Bayesian metrics and comprehensive scenario evaluation~\cite{ref2}.

\paragraph{Comparison of Stopping Boundaries}

Research comparing Bayesian posterior probability monitoring to frequentist group sequential designs (GSD) reveals interesting parallels.

\begin{table}[htbp]
\centering
\caption{High-level mapping from common sequential design paradigms to typical efficacy stopping rules and key design inputs (inspired by Table 2 in~\cite{ref4}).}
\label{tab:stopping}
\begin{tabular}{@{}p{3.2cm}p{6.0cm}p{5.0cm}@{}}
\toprule
\textbf{Paradigm} & \textbf{Typical efficacy stopping rule} & \textbf{Key inputs to specify} \\
\midrule
\multicolumn{3}{@{}l}{\textbf{Bayesian designs:}}\\
\addlinespace
Posterior probability rule & Stop (or declare success) when the posterior probability of a clinically relevant effect exceeds a chosen cutoff & Prior for treatment effect; posterior-probability cutoffs at interim and final looks \\
\addlinespace
Predictive-probability monitoring & Stop when the posterior predictive probability of ultimate trial success is sufficiently high (efficacy) & Prior for treatment effect; final success rule; interim predictive-probability cutoffs \\
\addlinespace
Decision-theoretic design & Choose the action that minimizes posterior expected loss (e.g., stop for efficacy when expected utility favors approval) & Prior for treatment effect; loss/utility function and the set of allowable actions \\
\addlinespace
\multicolumn{3}{@{}l}{\textbf{Frequentist designs:}}\\
\addlinespace
Group sequential testing & Stop for efficacy when the interim test statistic crosses a pre-defined efficacy boundary & Test statistic; boundary shape/alpha-spending (critical region) across looks \\
\addlinespace
Stochastic curtailment & Stop when conditional power for final success is sufficiently high (or low for futility), given an assumed effect & Final critical value; assumed effect for CP; interim CP cutoffs \\
\bottomrule
\end{tabular}
\end{table}


\section{Bayesian Computational Frameworks and Software}

The practical feasibility of Bayesian clinical trials has been enabled by advances in computational power and the development of Markov chain Monte Carlo (MCMC) algorithms. MCMC allows for the sampling from complex posterior distributions that cannot be solved analytically~\cite{ref9, ref45}.

\subsection{Technical Benchmarks and Toolsets}

The software landscape for Bayesian clinical trials is primarily centered on the R statistical environment, with specialized interfaces for modern Bayesian computation and model checking~\cite{ref46, ref45}.

\begin{table}[htbp]
\centering
\caption{Technical Benchmarks and Toolsets}
\label{tab:software}
\begin{tabular}{@{}p{2cm}p{3cm}p{4.5cm}p{4.5cm}@{}}
\toprule
\textbf{Software} & \textbf{Computational Method} & \textbf{Target Use Case} & \textbf{Performance Metric} \\
\midrule
Stan (rstan/brms) & Hamiltonian Monte Carlo (NUTS) & Complex hierarchical models; high-dimensional parameters & Robust performance for challenging posterior geometries in many applications~\cite{ref45} \\
\addlinespace
JAGS/rjags & Gibbs Sampling & Standard conjugate models; simple regression & Can be less efficient than HMC for complex high-dimensional posteriors~\cite{ref45, ref47} \\
\addlinespace
INLA & Laplace Approximation & Latent Gaussian Models; rapid interim analysis & 26$\times$ to 1800$\times$ faster than MCMC; often yields close agreement with MCMC for many latent Gaussian models (approximation, not a general guarantee)~\cite{ref47} \\
\addlinespace
Nimble & Customizable MCMC & User-defined samplers; specialized design modeling & Flexible; enables custom samplers and model code generation~\cite{ref48} \\
\bottomrule
\end{tabular}
\end{table}

Integrated Nested Laplace Approximations (INLA) have emerged as a fast alternative to simulation-based MCMC for latent Gaussian models. In recent benchmarks using data from a COVID-19 adaptive platform trial, INLA was between 85 and 269 times faster than Stan while showing close agreement in marginal uncertainty summaries (e.g., 95\% credible intervals) in that setting~\cite{ref47}. As an approximation method, INLA's accuracy is problem-dependent; nonetheless, its computational speed can be attractive for interim monitoring when rapid turnaround is needed~\cite{ref47}.

\section{Expert Prior Elicitation: Quantifying Subjectivity}

In scenarios such as ultra-rare diseases or novel therapeutic mechanisms, there may be limited historical data to support informative priors. In these instances, Bayesian analyses may rely on structured expert elicitation, with careful bias mitigation and transparent documentation of elicitation procedures and resulting distributions~\cite{ref49, ref50}.

\subsection{Structured Elicitation Protocols}

Elicited priors are prone to cognitive biases (e.g., overconfidence or anchoring). Structured elicitation frameworks aim to transform expert judgments into distributions while mitigating bias and enabling sensitivity analyses~\cite{ref49, ref50}.

\begin{itemize}[leftmargin=*]
    \item \textbf{SHELF (Sheffield Elicitation Framework):} A group-based protocol where experts provide probability judgments on treatment effects. It uses a facilitator to guide behavioral aggregation and reach a group consensus distribution~\cite{ref49, ref50}.

    \item \textbf{Delphi-style aggregation:} Iterative elicitation where experts provide judgments, review anonymized summaries, and refine assessments over multiple rounds to improve calibration and reduce anchoring~\cite{ref49, ref50}.

    \item \textbf{MATCH:} A user-friendly web interface for the SHELF methodology, allowing remote elicitation and providing real-time visual feedback to the experts regarding the implications of their probability assignments~\cite{ref49}.
\end{itemize}

In practice, elicitation often involves training experts, piloting with calibration questions, documenting aggregation choices, and evaluating robustness of trial conclusions to alternative elicited priors~\cite{ref50, ref2}.

\section{Practical submission checklist}

For submissions relying on Bayesian methods for primary inference, the FDA draft guidance emphasizes that reviewers should be able to understand, reproduce, and stress-test the design and analysis~\cite{ref2}. The items below correspond to major guidance sections on \emph{Success Criteria and Operating Characteristics}, \emph{Prior Distributions} (including borrowing and sensitivity analyses), and \emph{Software and Computation}~\cite{ref2}. A practical checklist includes:

\begin{itemize}[leftmargin=*]
  \item \textbf{Success criteria:} Clear estimands and success thresholds (e.g., posterior-probability criteria), with clinical justification and an explanation of whether criteria are calibrated to frequentist Type I error rate or to alternative Bayesian operating metrics with sponsor--FDA agreement~\cite{ref2}.
  \item \textbf{Priors and borrowing:} Rationale for each prior (analysis prior and any design/sampling priors), relevance of external data, discounting/robustness strategies, and sensitivity analyses addressing prior--data conflict~\cite{ref2}.
  \item \textbf{Operating characteristics:} Scenario set, simulation design, and summaries (e.g., Type I error rate/power when calibrated; Bayesian power/assurance and probability of correct decision when not calibrated), including how assumptions were chosen and what constitutes unacceptable performance~\cite{ref2, ref4}.
  \item \textbf{Computation and reproducibility:} Software/version details, documented code, key algorithm settings (including MCMC settings and reported seed number(s) when applicable), and diagnostics supporting convergence/reliability~\cite{ref2}.
\end{itemize}

\section{Discussion, limitations, and conclusion}

The 2026 FDA draft guidance makes Bayesian methods a practical regulatory topic rather than a purely methodological one: when used for primary inference, Bayesian designs must be specified so that reviewers can evaluate whether conclusions are robust to clinically plausible uncertainties in priors, models, and data quality~\cite{ref2}. This emphasis is particularly relevant for settings that motivate Bayesian designs in the first place (e.g., rare diseases, pediatrics, platform trials, and borrowing external information), where data are limited and the consequences of prior misspecification or unrecognized bias can be large~\cite{ref2, ref23}.

The transition to broader Bayesian adoption is not without risk. The potential for prior misspecification---where a flawed prior biases inference or decision-making---remains a primary concern for both sponsors and regulators~\cite{ref2}. This motivates the routine use of skeptical/robust priors, prior--data-conflict diagnostics, and sensitivity analyses that transparently quantify how conclusions change under alternative priors and modeling assumptions~\cite{ref2, ref7}. In addition, Bayesian designs can increase operational complexity (e.g., real-time data flow, simulation infrastructure, and computational monitoring), which can become a practical failure mode if not planned and resourced appropriately~\cite{ref2, ref9}.

Finally, it is important to recognize limitations of this perspective. First, the FDA document is a \emph{draft} guidance and may evolve; sponsors should engage early with FDA on design choices and evidentiary standards for specific programs~\cite{ref2, ref3}. Second, this paper emphasizes high-level expectations rather than providing a full technical manual for every Bayesian design family. In practice, regulatory acceptability depends on whether the pre-specified estimand, model, prior, and decision criteria together yield acceptable operating characteristics under scenarios that reflect key clinical and operational uncertainties~\cite{ref2}.

In conclusion, FDA's 2026 draft guidance can be read as a two-part message: Bayesian analyses can support primary inference in drug and biologic trials, and such analyses must be accompanied by explicit success criteria, principled and stress-tested priors (especially for borrowing), prospective operating-characteristic evaluation, and reproducible computation suitable for regulatory verification~\cite{ref2}. Sponsors who treat these items as core design deliverables---not post-hoc documentation---will be best positioned to deploy Bayesian methods credibly in regulatory submissions.

\end{document}